# Modeling the Behavior of Complex Aqueous Electrolytes Using Machine Learning Interatomic Potentials: The Case of Sodium Sulfate

Ademola Soyemi and Tibor Szilvási*

Department of Chemical and Biological Engineering, The University of Alabama, Tuscaloosa, Alabama 35487, United States

* Email: tibor.szilvasi@ua.edu

**Abstract**

Understanding the structure and thermodynamics of solvated ions is essential for advancing applications in electrochemistry, water treatment, and energy storage. While ab initio molecular dynamics methods are highly accurate, they are limited by short accessible time and length scales whereas classical force fields struggle with accuracy. Herein, we explore the structure and thermodynamics of complex monovalent-divalent ion pairs using $Na_2SO_4$(aq) as a case study by applying a machine learning interatomic potential (MLIP) trained on density functional theory (DFT) data. Our MLIP-based approach reproduces key bulk properties such as density and radial distribution functions of water. We provide the hydration structure of the sodium and sulfate ions in 0.1-2 M concentration range and the one-dimensional and two-dimensional potentials of mean force for the sodium–sulfate ion pairing at the low concentration limit (0.1 M) which are inaccessible to DFT. At low concentrations, the sulfate ion is strongly solvated, leading to the stabilization of solvent-separated ion pairs over contact ion pairs. Minimum energy pathway analysis revealed that coordinating two sodium ions with a sulfate ion is a multistep process whereby the sodium ions coordinate to the sulfate ion sequentially. We demonstrate that MLIPs allow the study of solvated ions beyond simple monovalent pairs with DFT-level accuracy in their low concentration limit (0.1 M) via statistically converged properties from ns long simulations.

**Introduction**

The structure and thermodynamics of solvated ions play an important role in understanding processes that relevant to electrochemistry, biochemistry, and water treatment among others.[1-3] For example, ion interactions in solution play a critical role in energy storage applications,[4] $CO_2$ capture,[5] $CO_2$ reduction,[6] and ion diffusion in biological membranes[7,8]. Additionally, gaining deeper insights into the solvation structure and behavior of ions near at a solid-liquid interphase is also crucial to understanding the activity and stability of electrocatalytic/electrochemical electrode materials, or membranes used in separation processes.[9-15] The performance of these technologies depend on the specific ion-ion and ion-solvent interactions and thus have important implications for developing improved separation and energy storage technologies.

In recent years, significant advances in computational methods have made *ab initio* molecular dynamics (AIMD) studies using density functional theory (DFT) more common in studying the dynamics and thermodynamic properties of electrolyte solutions such as the potential of mean force (PMF), chemical potential, solvation structure, and ion diffusion among others.[3, 16-21] In particular, the PMF between an ion pair in solution provides a direct window into the ion pairing behavior, solution structure of the electrolyte, and can help understand ion-ion, ion-solvent or ion-surface interactions. To understand ion pairing and their dynamics in solution, the PMF can be studied as a function of various collective variables such as ion-ion distance, ion coordination number, or ion-ion angle for divalent ion systems (2:1 electrolyte systems).[1, 3, 19] AIMD studies of the PMF are however limited despite the accuracy of DFT due to short time and length scales that are accessible to DFT. As a result, AIMD studies have mostly been limited to 1-dimensional (1D) PMFs of simple monovalent ions and/or single cation-anion pairs (or 1:1 electrolyte systems),[1, 3, 16, 22-24] while some practitioners have approximated the simulation of divalent ions (or 2:1 electrolyte systems) by considering only one anion-cation pair while completely ignoring the second counterion and balancing the total charge by utilizing a uniform background charge.[19, 25] The computational limitations of AIMD is clear for monovalent-divalent ion pairs (or 2:1 electrolyte systems) for which longer time and length scales may be required to properly converge 2-dimensional PMFs to get a comprehensive understanding of ion dynamics and behavior. Thus, only classical force fields have been employed for explicitly describing monovalent-divalent ion pair electrolyte solutions via PMF simulations.[26-30] Classical force fields are however inaccurate for describing complex interactions in aqueous solutions such as hydrogen bonding and polarizability even after careful tuning of force field parameters relative to experimental or DFT data.[31]

To circumvent the limitations of AIMD and classical force fields, machine learning interatomic potentials (MLIPs) offer a balance in cost and accuracy in simulating the PMF of ion pairs. MLIPs that are trained on DFT reference data have recently gained popularity because MLIPs can in principle be used to carry out DFT-quality simulations while gaining access to longer time and length scales.[3, 9, 16, 24, 32, 33] While the PMF of relatively simple systems such as metal halides like $LiCl_{(aq)}$ and $NaCl_{(aq)}$ have been studied using MLIPs,[3, 24, 28, 29, 34-36] understanding of more complex solutions with monovalent-divalent ion pairs is still lacking.

In this computational study, we show how MLIPs can describe complex neutral aqueous solutions with monovalent-divalent ion pairs using the example of $Na_2SO_4$. The hydrated sulfate ion ($SO_4^{2-}$) is present in a range of processes in chemistry and biology as they are abundant in many earth-abundant minerals,[27] in salt water as well as in freshwater environments,[37, 38] and also play a role in climate change due to sulfate aerosols in the atmosphere.[39] Sulfate ions have also been applied in electrolytes of sodium-ion batteries and have been shown to be kosmotropic in nature, and thus significantly alter the structure of water.[27, 40-42] Consequently, comprehensive understanding of the behavior of the sulfate ion in water and its interaction with common cations such as $Na^+$ is important for advancements in environmental science, industrial chemistry, and beyond. In this work, we reproduce bulk properties such as density and radial distribution functions (RDFs) and provide DFT-quality 1D and 2D PMFs for understanding the sodium sulfate ion pairing behavior. We also study the solvation structure of the ions and the impact of $Na_2SO_4$ concentration on water diffusivity, the hydrogen bonding and tetrahedrality of water. We find that our trained MLIP recovers the structural features (such as RDFs and density) of bulk water compared to DFT and in addition reproduces the solvation structure of the ions as well as the density vs concentration trends relative to experiments. We observe that $Na_2SO_4$ disorders the hydrogen bonding network and tetrahedrality of water. Additionally, the diffusivity of water, sodium and sulfate ions reduces with concentration in agreement with experiments. Our results also show that at the low concentration limit and in the 1-dimensional picture (i.e., taking only one sodium-sulfate distance into account), the solvent-separated ion pair (SSIP) is energetically more stable (~3 kJ/mol) than the contact-ion pair (CIP) due to the strong solvation of the sulfate ion and weak interaction with the sodium ion.[43, 44] The 2D PMF also shows that the sulfate ion prefers to be segregated from both sodium ions at the low concentration limit, implying that the electrostatic attraction between the ions is effectively screened by the solvating water molecules. Finally, we show the minimum energy pathway for coordinating two sodium ions with a sulfate ion is a multistep process whereby the sodium ions coordinate the sulfate ion one after the other. Beyond producing computational predictions of $Na_2SO_4$ properties in water, our results can also serve as a reference for future validation of MLIPs that include explicit treatment of long-range interactions,[45] such as long-range electrostatics whereby the influence of long-range interactions can be directly evaluated. This work also motivates future X-ray-based studies such as EXAFS and SAXS of $Na_2SO_{4,aq}$ solutions to further elucidate the short- and long-range structure of these ions in water. Additionally, explicitly characterizing the long-range structure and correlations (> 10 Å) in $Na_2SO_4$ electrolyte solutions using

MLIPs is beyond the scope of this study. As established by Kirkwood,[46] such analysis requires either large-scale molecular simulations,[47] or molecularly informed liquid-state theories,[48] capable of capturing the exponential or damped-oscillatory decay of charge–charge correlations. An additional future research avenue might be to assess universal MLIPs in the modeling of aqueous electrolytes. Our work and generated dataset can thus also serve as a useful starting point and benchmark for the fine-tuning and validation of universal MLIPs. Overall our study highlights that MLIPs allows the study of complex solvated ions with DFT-level accuracy in their low concentration limit (0.1 M) and enables the calculation of the statistically converged properties of electrolyte systems from ns long simulations.

**Computational Methods**

   1. **MLIP Training Methodology**

All reference DFT calculations were performed using the Vienna Ab initio Simulation Package (VASP) version 6.3.2.[49, 50] The exchange and correlation contribution of the electronic energy were computed using the RPBE functional.[51] The electron-core interaction was represented by projector augmented waves (PAWs) with a plane wave basis energy cutoff of 520 eV. In all single-point calculations, the Brillouin zone was sampled using a gamma k-point grid and the orbital energies were broadened using a Gaussian smearing width of 0.03 eV. To obtain accurate forces, the energy convergence criteria was set to $10^{-7}$ eV. Previous studies[52, 53] have shown that dispersion-corrected RPBE accurately describes bulk water and consequently we include dispersion corrections using Grimme's D3 method with zero damping.[54] In addition to this, a recent study has shown that the D3 method inaccurately calculates the $C_{AB}^6$ term for pairs involving cations such as $Na^+$.[55] Consequently, we modified the DFT-D3 code to exclude the contributions of $Na^+$.

Curation of a training set that sufficiently reproduces the potential energy surface (PES) is crucial to the construction of an accurate MLIP.[56-58] Since the PES does not have a physically derived functional form, the MLIP can only learn the correct PES shape if all relevant parts of the configuration space are sampled from. Specifically, all subsystems such as bulk water, sodium sulfate in water (at different concentrations initialized using Packmol[59]), must be well represented in the training set. To accomplish this, we carried out 0.5 ns NVT MD simulations at different densities for bulk systems using the MACE-MP-0 pretrained MLIP[60] via the Atomic Simulation Environment (ASE, version 3.22.1).[61] These simulations were run with deuterated hydrogens at 300 K under Langevin dynamics and a 1 fs timestep. All MD simulations of bulk water were run using 128 water molecules and 0 to 5 $Na_2SO_4$ ions to simulate a concentration range of 0 to 2 M. The densities of the simulation cells were also varied with in ±4% of experimental densities.[62] Additionally, since NVT simulations typically provide equilibrium configurations we carried out constant potential energy MD simulations (also known as contour exploration)[63] whereby we explicitly generate high energy structures which have been shown to improve the accuracy and stability of the resulting MLIP.[56] Here, we perform constant potential energy MD simulations, using the bulk water systems (with and without ions), with a max step size of 1 Å for 10000 steps at each energy level up to 5 eV above the energy of the optimized geometry with an energy step size of 0.5 eV. Single point energy calculations were performed on uncorrelated structures chosen at 3 ps intervals and were then used as the initial training and validation set.

We train MLIPs using NequIP package.[64] The development branch of NequIP downloaded on October 23, 2023, from GitHub (https://github.com/mir-group/nequip) is used for all training and inference. The initial dataset generated as described above comprised of 1000 structures. This was split into training set (80%) and validation set (20%) and multiple NequIP [64] MLIPs with a 6 Å radial cutoff and equivariant E(3) products up to $L_{max}$ = 2 in the tensor layers were trained to find a MLIP with the optimal set of hyperparameters. The NequIP MLIPs were trained in two stages of 500 epochs each – first where the force to energy ratio in the loss function is 20:10, ensuring the forces are trained accurately, and second where the force to energy ratio is 10:10. In both stages the stress loss coefficient is set high (100,000) to ensure the stresses are also sufficiently learned. Eight of the best MLIPs based on the validation loss we then selected to be used as the committee of MLIPs (See Supporting information for training input files for all committee members). Training and inference of our MLIPs were performed using an NVIDIA MSI GeForce RTX 4090 GPU, and 2 AMD EPYC 7551P CPU cores.

For generating new structuress to be added to the initial training set, NequIP-based MD simulations were performed using the OpenMM-ML plugin (version 1.1) in the OpenMM[65] code (version 8.1.1) except for contour exploration simulations which were performed in ASE. In addition to this, we also utilized umbrella sampling to generate specific configurations of ions in bulk water (at 0.4 M concentration). All umbrella sampling MD simulations were performed using OpenMM in which the PLUMED[66-68] plugin (version 2.0.1) in OpenMM was used to apply harmonic umbrella potentials of the from $V(r) = k(r_0 - r)^2$ with a force constant k of 200 kJ/mol/nm$^2$ on the collective variable (i.e., sodium-sulphate distance or sodium-sodium distance). To select new structures that were added to the original training set, we followed a query-by-committee approach whereby critical structures are selected based on uncertainty of the committee of MLIPs in the energy, force predictions (see Supporting Information for more details). In each iteration, we add only 100 to 300 structures to the training and validation sets, respectively. The active learning process is repeated until an MLIP that is stable and sufficiently reproduces bulk properties is obtained. We find an approximately linear correlation between the committee uncertainty and true error (Figure S14), indicating that uncertainty-based selection is suitable for selecting new structures without calculating the ground truth.[56, 69] Our converged MLIPs required 5 iterations with a final training set of 1849 structures and a validation set of 1000 structures while covering all relevant distances and bond lengths in this system (See Figures S8 to S13). Finally, to generate an independent test set we ran 500 ps NPT simulations with our final MLIP using 128 water molecules and 0 to 5 Na$_2$SO$_4$ (0 to 2 M concentration) using the same setup described above. To ensure that selected structures are uncorrelated, we selected every 400$^{th}$ structure from each MD and ended up with 12,820 structures which were then labeled at DFT level and used as an independent test set to evaluate the accuracy of our MLIP. The final MLIP energy MAE is trained to 0.29 meV/atom, the force MAE is trained to 14 meV/Å, and the stress MAE is trained to 0.06 meV/Å$^2$ (See Supporting Information for correlation and distribution plots of errors).

2. **MLIP-based MD Methodology**

For production runs to obtain structural and dynamic properties such as density, radial distribution functions (RDFs), tetrahedrality, and hydrogen bonding dynamics and so on, we carried out MD simulations in simulation boxes containing 256 water molecules to minimize finite size effects and with 0 to 10 Na$_2$SO$_4$ molecules (0 – 2M concentration). While for a 0.1 M concentration, we had 1 Na$_2$SO$_4$ molecule in 512 water molecules. Initial structures were obtained using Packmol,[59] and the system was optimized and then equilibrated within the NVT ensemble under Langevin dynamics and a 0.5 fs timestep at 300 K for 50 ps using OpenMM. Following equilibration, NPT simulations were then run whereby we utilized OpenMM's MonteCarlo barostat to maintain pressure at 1 atm and run at 300 K for 4 ns using a timestep of 0.5 fs. The trajectory was saved to file every 50 fs. For RDF analysis we utilized the TRAVIS code,[70, 71] and studied the water O-O, water O-H, water H-H, sodium-water (Na-Ow), sulfate-water (S-Ow), sulfate-sodium (S-Na), sulfate O – water H (Os-H), and sulfate – water H (S-H) RDFs. Since NPT simulations are inappropriate to obtain diffusivities,[33] we performed NVT simulations using 256 water molecules following the same procedure described above except for the 0.1 M concentration where 512 water molecules was used. Here, the volume of the simulation box was chosen to match the MLIP-predicted equilibrium volume. The diffusivity of water and hydrogen-bonding network was computed using MDAnalysis (version 2.8.0).[72] To compute the number of hydrogen bonds per water molecule, we define an intact H-bond when the O-O distance $R_{OO}$ < 3 Å and the $O_D$-$H_D$---$O_A$ angle formed by the donor oxygen $O_D$, donor hydrogen $H_D$ and the acceptor oxygen $O_A$ is greater than 150°. To compute the oxygen-oxygen-oxygen triplet angle distribution ($P_{OOO}(\theta)$), three O atoms were considered as part of a given triplet if two of the water molecules (i.e., O atoms) were within a cutoff distance from the chosen central water molecule. This cutoff distance was chosen as 3.4 Å so that the average O–O coordination number is approximately 4 at the cutoff distance.[73, 74] Meanwhile the tetrahedral order parameter, which is defined as $q_t = 1 - \frac{3}{8}\sum_{i=1}^{3}\sum_{j=i+1}^{4}(\cos(\theta_{ij}) + \frac{1}{3})^2$ where $\theta_{ij}$ is the angle formed by a given central water molecule and its nearest neighbors i and j, was computed using the ORDER code (version 0.0.3).[72, 75] To gain insight into the denticity of $Na^+$ ions with $SO_4^{2-}$ ions, we also computed the S-O-Na angle. In this analysis, we defined a coordinated $Na^+$ when the S-Na distance is < 3.8 Å and the Na-O distance is < 3 Å

3. **MLIP-based Umbrella Sampling MD Methodology**

To obtain $Na^+$ – $SO_4^{2-}$ PMFs, we carried out umbrella sampling simulations in simulation boxes containing 1 $Na_2SO_4$ and 512 water molecules, i.e., at 0.1 M concentration (See Figure 5 for convergence of PMF with system size). After packing the simulation box using Packmol,[59] the system was optimized and equilibrated within the NVT ensemble under Langevin dynamics and a 0.5 fs timestep at 300 K for 50 ps. Following equilibration, NPT simulations were then run whereby we utilized OpenMM's MonteCarlo barostat to maintain pressure at 1 atm and run at 300 K for 4 ns using a timestep of 0.5 fs which was sufficient to obtain a well converged PMF (Figure S15). Umbrella sampling windows for the $Na^+$ – $SO_4^{2-}$ distance ranging 3 to 9 Å were equally spaced by 1 Å employing harmonic umbrella potentials of the from *V(r) = k(r₀ – r)²* with a force constant k of 200 kJ/mol/nm². In addition to this, we included additional umbrella sampling windows using a 6000 kJ/mol/nm² harmonic restraining force/bias in the barrier region that exists between the solvent-separated ion pair (SSIP) and contact ion pair (CIP) states (~4.5 Å), and for distances beyond the radial cutoff of the MLIP (i.e. > 6 Å) to compensate for the finite radial cutoff of the MLIP (6 Å). In each simulation, the value of the $Na^+$ – $SO_4^{2-}$ distance was saved every 50 steps. The weighted histogram analysis method (WHAM)[76] was employed to extract a free-energy profile from these histograms. To obtain the PMF from the WHAM free energy, a $2k_BT \cdot \ln(x)$ term is added to the free energy, where x and T represent the $Na^+$ – $SO_4^{2-}$ distance and temperature, respectively. Additionally, the uncertainty of the PMF was obtained via the bootstrapping algorithm implemented in the WHAM code (version 2.0.11). For the 2D PMF, we followed the same procedure as above but, in this case, the restraining bias was applied to both sodium-sulphate distances. To obtain 2D the PMF from the WHAM free energy, a $2k_BT \cdot \ln(x) + 2k_BT \cdot \ln(y)$ term is added to the free energy, where x and y represents the $Na_1^+$ – $SO_4^{2-}$ and $Na_2^+$ – $SO_4^{2-}$ distances, respectively, while T represents temperature. Lastly, minimum energy pathway analysis was performed using the MESPA code (version 1.4).[77]

4. **AIMD Methodology**

To validate our prediction of bulk water density and structural properties and show the reliability of our MLIP, we perform a 60 ps AIMD simulation of bulk water using 128 water molecules. The AIMD simulation was performed in the Vienna Ab initio Simulation Package (VASP) version 6.3.2.[49, 50] The exchange and correlation contribution of the electronic energy were computed using the RPBE functional[51] and dispersion corrections were accounted for using Grimme's D3 method with zero damping.[54] The electron-core interaction was represented by projector augmented waves (PAWs) with a plane wave basis energy cutoff of 520 eV and the Brillouin zone was sampled using a gamma k-point grid and the orbital energies were broadened using a Gaussian smearing width of 0.03 eV. To obtain accurate forces, the energy convergence criteria was set to $10^{-7}$ eV. The simulation was run in the NPT ensemble with a timestep of 0.5 fs and at a temperature of 300 K, with a friction factor of 10 ps$^{-1}$ for the Langevin thermostat. For the density analysis and generation of RDFs, we discard the first 10 ps of the simulation and selected every fifth frame from the remaining 50 ps.

**Results and Discussion**

We have computed the structural and dynamic properties of $Na_2SO_4$ in water at varying concentrations using our MLIP and compared our MLIP predictions to DFT and experimental data where possible. By comparing our prediction of these properties to DFT and experiments we establish the accuracy of our MLIP. We begin our analysis by demonstrating the accuracy of our MLIP in reproducing experimental density trends with varying concentrations of $Na_2SO_4$ in water, and DFT-predicted structural properties of pure bulk water (See Supporting Information for analysis of energy, force, and stress error metrics). We then discuss the structural properties of the ions in water such as the ion-water RDFs, ion-ion RDF and show that our MLIP recovers experimentally observed peak positions and coordination numbers. Next, we discuss the tetrahedrality, hydrogen bonding, and transport properties of water and the effect of $Na_2SO_4$ on these properties. Lastly, we discuss the free energy landscape of $Na_2SO_4$ in water through the 1D and 2D PMF.

Figure 1 (top left) shows the predicted variation of density with increasing $Na_2SO_4$ concentration. Here, we see that the MLIP captures the trend in density with $Na_2SO_4$ concentration and relative errors range from -0.5 and -1.2% as shown in Table S2. We also note the good agreement with the RPBE-D3(0) predicted bulk water density (MLIP water density of 0.995 g/cm$^3$ vs RPBE-D3(0) bulk water density of 1.008 g/cm$^3$). While recent studies[78, 79] have shown RPBE-D3 bulk water density to be ~ 0.90 g/cm$^3$ using MLIPs, we highlight that this discrepancy can be due to several factors such as differences in training set, active learning approach, or application of external pressure. Figure 1 (top right) compares the O-O RDF in bulk water predicted by the MLIP to the DFT ground truth. Here, we see excellent agreement between our MLIP and DFT even for distances beyond the radial cutoff of the MLIP (pink background) that lies outside the radial cutoff of the MLIP thus showing that the performance of the MLIP does not degrade beyond its radial cutoff. The good agreement beyond the MLIP's radial cutoff also shows that bulk water structure can be recovered without explicit treatment of interactions beyond the radial cutoff of the model (6 Å) due to screening effects.[80] Table S3 shows that for the O-O RDF, the MLIP reproduces the water structure accurately with an average absolute error of 1.0 % with a range of 0 to 2.5 % error in the RDF features. Figure 1 (bottom left) shows the comparison of the O-H RDF between the MLIP and DFT. Similarly, our MLIP accurately reproduces the peak positions and shape of the DFT O-H RDF even at radial distances beyond the MLIP's radial cutoff, and as shown in Table S3, there is an absolute average absolute error in the RDF features of 1.8 % with a range of 0 to 5.3 % error. Figure 1 (bottom right) shows the comparison of the H-H RDF between the MLIP and DFT. Likewise, our MLIP accurately reproduces the peak positions and shape of the DFT H-H RDF even at radial distances outside the MLIP's radial cutoff, and as shown in Table S3, there is an absolute average absolute error in the RDF features of 1.5 % with a range of 0 to 2.9 % error. The high accuracy of our MLIP in predicting the RDFs, and density trends thus highlights the reliability of our MLIP in modeling $Na_2SO_4$ in water.

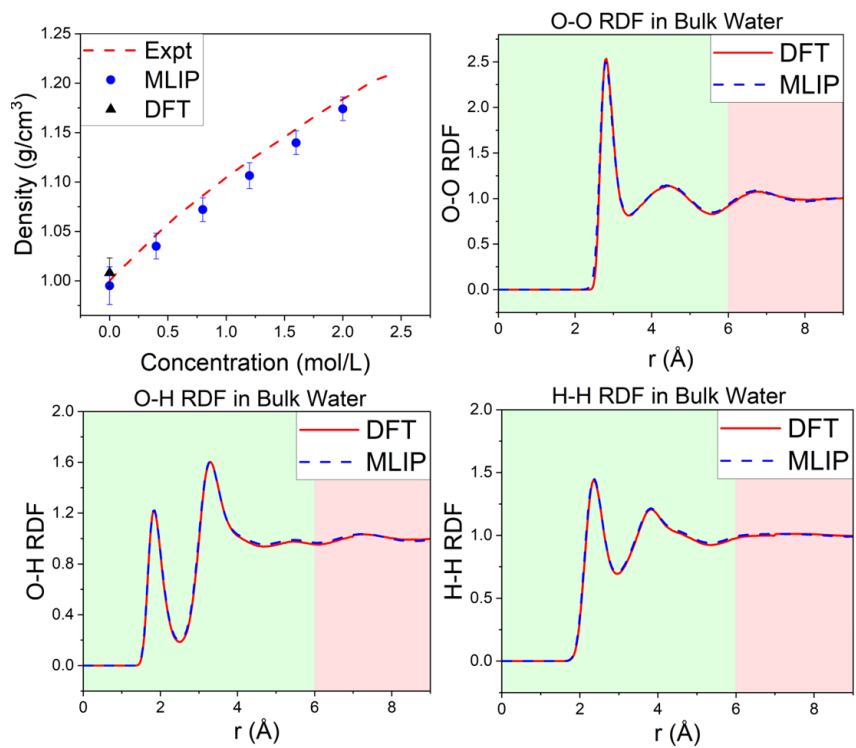

**Figure 1**. Variation of the density with increasing $Na_2SO_4$ concentration compared to experimental values[62] (top left). Comparison between MLIP-predicted and DFT-predicted water O-O (top right), O-H (bottom left), H-H (bottom right) RDF. The green and pink background indicates radial distances that fall within and outside the radial cutoff of our MLIP (6 Å), respectively. DFT RDF data was obtained from a 60 ps AIMD simulation. DFT simulations were performed at RPBE-D3(0)/520eV level of theory in VASP. Note that the cell was doubled in all directions in TRAVIS to obtain the RDF beyond 7 Å for the AIMD data which contained only 128 water molecules.

Next, we analyze the $Na^+$ and $SO_4^{2-}$ hydration structure in water at 0.1 M concentration. The hydration structure of the $Na^+$ ion has been extensively studied experimentally through various methods such as X-ray diffraction, neutron diffraction, and extended X-ray absorption fine structure methods.[22, 23, 26, 81, 82] These studies have shown the Na-water 'bond length' to fall within the 2.38 and 2.50 Å range and is independent of concentration. Furthermore, the CN has been estimated to be around 6. Figure 2 (top left) shows the Na-Ow RDF predicted by our MLIP. Compared to the experimental first Na-Ow peak and CN, the MLIP provides good agreement with a predicted peak position of 2.42 Å, and a CN of 5.34 (see Table 1 and Table S4). On the other hand, the $SO_4^{2-}$ hydration structure has been less studied mostly due to the difficulty of the experiments and difficultly in correctly separating the contributions of the sulfate and water oxygen atoms.[26, 27, 82-84] Despite this, experimental studies have estimated the S-Ow peak to fall in the 3.67 to 3.89 Å range, while the CN is in the 6.0 to 14.3 range (see Table 1 and Table S4). It should again be noted that the wide range of values for the CN values is attributed to the difficulty of performing the experimental measurements. Figure 2 (top right) shows the S-Ow RDF predicted by our MLIP and compared to the experimental first S-Ow peak and CN, the MLIP provides good agreement with a predicted peak position of 3.75 Å, and a CN of 11.10 which falls within the experimental range of values. Furthermore, in agreement with previous experimental results, the hydration structure of sodium and sulfate ions does not change significantly with concentration as shown in Figure 2 (bottom left and bottom right).

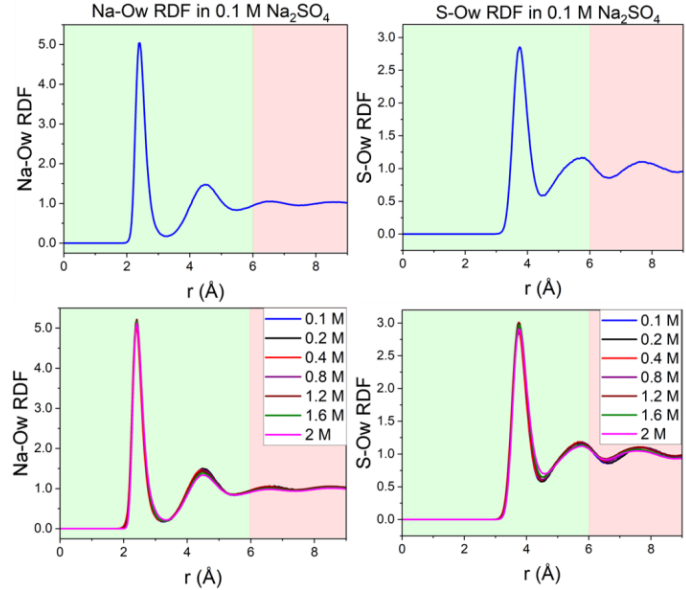

**Figure 2**. Na-Ow RDF (top left) at 0.1 M concentration, S-Ow RDF (top right) at 0.1 M concentration, Na-Ow RDF (bottom left) between 0.1 and 2 M concentration, and S-Ow RDF (bottom right) between 0.1 and 2 M concentration. The green and pink background indicates radial distances that fall within and outside the radial cutoff of our MLIP (6 Å), respectively. Note that the y-axis on plots on the Na-Ow and S-Ow plots are on different scales.

**Table 1**. Comparison of ion hydration structure from MLIP-based MD (at 0.1 M concentration) against experiments. $r^1_{max}$, $g^1_{max}$, $r^1_{min}$, $g^1_{min}$, and CN refer to the position of the first peak in the RDF, the height of the RDF at the first peak, position of the first minimum in the RDF, the height of the RDF at the first minimum, and the coordination number, respectively. CN is given by $CN = 4\pi\rho \int_0^{r^1_{min}} g(r)r^2 dr$, where $\rho$ is the number density, $r^1_{min}$, is the radial distance of first minimum in the RDF, and r is the radial distance from the reference atom. See Table S4 for the RDF features between 0.1 M and 2 M concentration.

|  | $r^1_{max}$ | $g^1_{max}$ | $r^1_{min}$ | $g^1_{min}$ | CN |
|---|---|---|---|---|---|
| Na - Ow | 2.42 | 5.03 | 3.22 | 0.17 | 5.34 |
| Exp[22, 23, 26, 81, 82] | 2.38-2.50 | - | - | - | 6 |
| S - Ow | 3.75 | 2.86 | 4.45 | 0.59 | 11.10 |
| Exp[26, 27, 82-84] | 3.67-3.89 | - | - | - | 6-14.3 |

To give further insights into the organization of the ions in water, we also studied the sulfate-sodium (S-Na) RDF at 0.1 M concentration as shown in Figure 3 (left). The first peak at 3.60 Å corresponds to the CIP state where the sulfate is in direct contact with one or two Na ions, which is in agreement with previous experimental studies that showed a peak position of 3.45 Å[85], while the second peak at 5.22 Å corresponds to the SSIP state where the ions are within their respective water solvation shells. It is also worth noting that the most important features of the sulfate-sodium RDF (i.e. the CIP and SSIP states) fall within the radial cutoff of the MLIP and are thus expected to be well-described. In Figure 3 (right), we find that the height of the first peak is relatively unchanged while the height of the second peak reduces as the concentration increases, suggesting that the SSIP state is destabilized relative to the CIP state (See Table S4 for all peak positions, heights and CNs). Thus, the relative abundance of CIPs increases with increasing concentration. The relative heights of the CIP and SSIP peaks and the CN of 0.03 suggest that the SSIP is primary form of ion association, and highlights that the ions are significantly stabilized by their respective hydration shells. This is also in agreement with previous experimental studies that have suggested that the electrostatic attraction between the ions is weakened in solution (particularly at low concentrations) leading to prevalence of SSIPs.[85, 86] Leaist and Goldik[86] also showed that even at a concentration of 2.25 M, the extent of association between the ions is 0.45 which is comparable to the MLIP predicted CN of 0.49 at 2 M (Table S4). Furthermore, due to the relatively low charge density of Na$^+$, it coordinates with the SO$_4^{2-}$ ion in monodentate form as shown in the calculated S-O-Na angle of 120° across the studied concentrations. This is in agreement with experimental X-ray studies which showed a S-O-Na angle of ~127°,[85] and further suggests that the ions do not interact strongly in solution.[86] In contrast, higher charge-density multivalent cations (e.g. Mg$^{2+}$, Ca$^{2+}$) can shed more of their solvation shell and simultaneously coordinate two oxygens on the same SO$_4^{2-}$, favoring bidentate coordination. The predominance of solvent-shared ion pairs (SSIPs) in aqueous Na$_2$SO$_4$ solutions would be useful in electrochemical applications as this indicates high solubility of the ions since direct ion contact is reduced, thereby leading to greater ion mobility and contributing to increased ionic conductivity. Moreover, high ion solubility could also be useful for applications of super-concentrated electrolytes in energy storage.[87] We also emphasize here that we expect our MLIP to maintain an accurate description for distances beyond the radial cutoff of the MLIP (> 6 Å) in the ion-water and ion-ion RDFs due to the message-passing nature of NequIP, and screening effects in bulk systems as has been observed in other studies involving electrolytes.[80, 88, 89] However, we suggest that future experimental studies should explore SAXS studies to give more insights into the structure of the Na$^+$ and SO$_4^{2-}$ ions in water and allow for more direct validation of MLIPs.

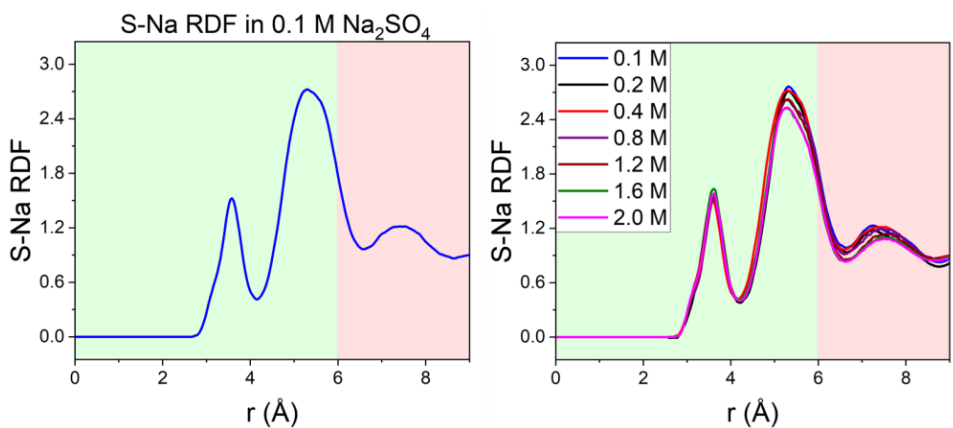

**Figure 3**. S – Na RDF at 0.1 M concentration (left), and S – Na RDF between 0.1 and 2 M concentration. The green and pink background indicates radial distances that fall within and outside the radial cutoff of our MLIP (6 Å), respectively.

In order to further analyze the local arrangement of water molecules in bulk water and in the presence of $Na_2SO_4$, we computed the distribution of oxygen-oxygen-oxygen triplet angles ($P_{OOO}(\theta)$) within the first coordination shell (see Figure 4 below), the tetrahedral order parameter $q_t$ (see Table S5), and the number of hydrogen bonds (HBs) per water molecule (see Table S5). In Figure 4 (left), we see that the MLIP accurately recovers the local tetrahedral network in bulk water compared to DFT and experiment as shown in the agreement between the MLIP predicted $P_{OOO}(\theta)$, DFT and experimental $P_{OOO}(\theta)$. Here, we observe that in general the MLIP (and DFT) underpredicts the probability of angles above 120° while overpredicting the probability of angles below 100°. Overall, the MLIP recovers the experimental peak at 100° which represents highly ordered ice-like water configurations. Meanwhile the shoulder that represents highly distorted hydrogen bond network is shifted by ~3° compared to experiments. Similarly, the $q_t$ value is predicted as 0.685 by our MLIP which is in relatively good agreement with the experimental value of 0.576.[73] Additionally, using our definition of HBs, we predict the average number of HBs per water molecule as 3.70.

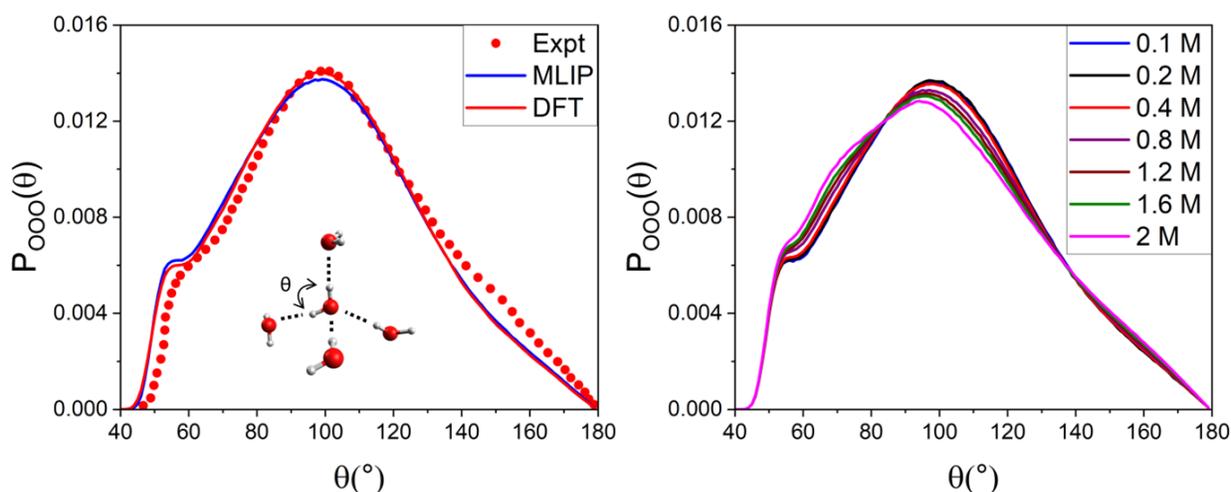

**Figure 4**. MLIP-predicted O-O-O triplet angular distribution functions ($P_{OOO}(\theta)$) of bulk water compared to DFT and experimental data[90] (left), and MLIP-predicted O-O-O triple angular distribution functions ($P_{OOO}(\theta)$) of water as a function of $Na_2SO_4$ concentration (right). The inset shows an illustration of the O-O-O triplet angle where the red and white spheres represent oxygen and hydrogen, respectively.

In Figure 4 (right), we show the effect of increasing $Na_2SO_4$ ion concentration on the tetrahedrality of water. Here, we see that as concentration increases from 0.1 M to 2 M the main peak position reduces from 98° to 95°. In addition, the shoulder at 55° is slightly intensified as concentration increases indicating that increasing $Na_2SO_4$ concentration subtly breaks the tetrahedrality of water and increases disorder in the hydrogen bonding network. This decrease in tetrahedrality is shown in the decrease in $q_t$ from 0.685 at 0 M concentration to 0.673 at 2 M concentration (see Table S5). Additionally, as the concentration increases, we observe the appearance of a new shoulder around 70° indicating formation of clusters with intermediate order. Furthermore, we find that the $N_{HB}$ also decreases with increasing $Na_2SO_4$ concentration as shown in Table S5 from 3.70 at 0 M concentration to 3.58 at 2 M concentration. We emphasize that since the notion of an intact or broken HB is somewhat ambiguous, we are only interested in the trends of the HBs as the concentration of $Na_2SO_4$ increases.

Next, we calculate the diffusivity/self-diffusion coefficient (D) of water as a function of the $Na_2SO_4$ concentration. As has been noted in previous studies the calculation of diffusivity can be greatly affected by finite size effects even with a simulation box containing 256 water molecules.[33, 74] Consequently, we analyze the effect of increasing $Na_2SO_4$ concentration based on the relative diffusivity of water molecules

($D/D_0$), i.e., the ratio of D at a given concentration to D in neat bulk water, as suggested in recent studies.[33] Table S6 shows the MLIP predicted diffusivity, $D/D_0$, and experimental $D/D_0$ values.[91] Here, we find that our MLIP is in qualitative agreement with experiments as $D/D_0$ monotonically decreases as concentration increases. This result further reinforces the notion that $Na_2SO_4$ is kosmotropic in nature since the strong interaction of the ions with water tends to reduce their motion.[92, 93] In addition to this, we also computed the diffusivities of the sodium and sulfate ions as a function of concentration as shown in Table S7. We find that in general, sodium ions are ~30% more mobile than the sulfate ions and that the diffusivity of the ions decreases as the concentration increases as a consequence of the increasing density of the solution.

Building on the insights gained from the RDF, the PMF offers additional understanding of the thermodynamic stability and free energy landscape governing sulfate-sodium interactions in solution. First, we tested the convergence of the 1D S-Na PMF with simulation length. As shown in Figure S15 we see that the PMF is converged after a 4 ns simulation compared to an 8 ns simulation, therefore, for all enhanced sampling simulations, a simulation length of 4 ns was used. Due to the relatively low barrier between the CIP and SSIP states, we also explored the ability of unbiased simulations to produce a well-converged S-Na PMF which also serves as a validation of our umbrella sampling simulation protocol. To obtain the S-Na PMF, we ran five parallel 8 ns NPT simulations at 0.1 M concentration whereby unique initial configurations were taken from the enhanced sampling simulations while taking the distance between the same S-Na pair as in the enhanced sampling simulations every 50 steps. Figure 5 (left) compares the S-Na PMF obtained from umbrella sampling (i.e. biased) simulations compared to the S-Na PMF obtained from unbiased simulations. Here, we can see that there is good agreement for close distances within the radial cutoff of the MLIP (green background) meanwhile there is some disagreement at larger separations beyond the radial cutoff (pink background). However, we note that the unbiased PMF falls within the uncertainty of 0.2 kJ/mol of the PMF obtained from biased simulations, thus we conclude that our umbrella sampling protocol appropriately captures the configurational space (See Figure S16).

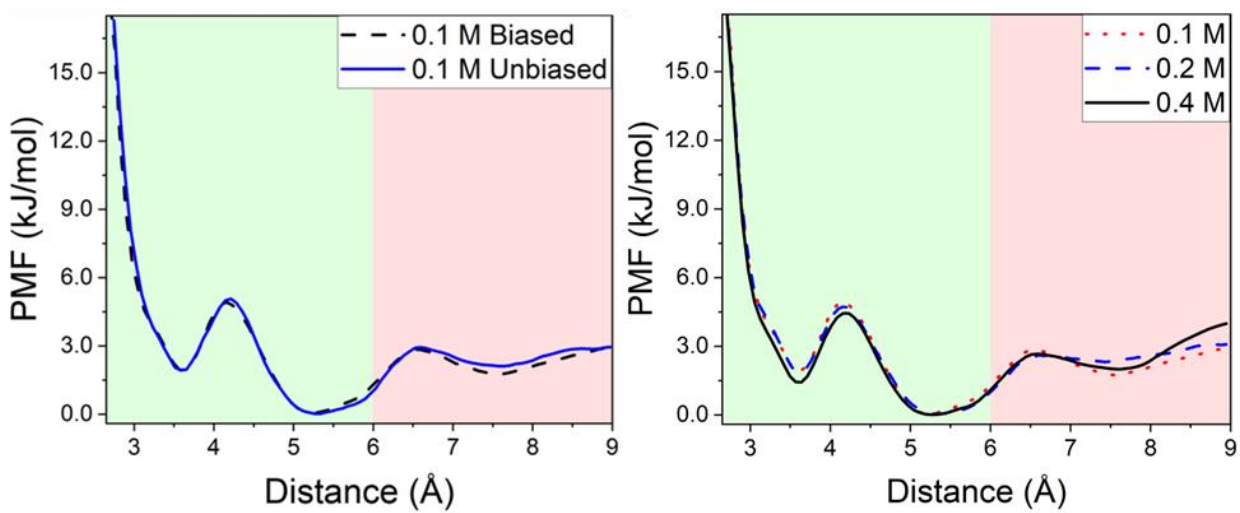

**Figure 5**. Comparison between S – Na PMF obtained from biased and unbiased MD simulations, respectively, at 0.1 M concentration (left), and S – Na PMF as function of $Na_2SO_4$ concentration (right). The green and pink background indicates distances that fall within and outside the radial cutoff of our MLIP (6 Å), respectively. Simulations were performed with one $Na_2SO_4$ and 128, 256, and 512 water molecules corresponding to 0.4 M, 0.2 M, and 0.1 M concentration of $Na_2SO_4$, respectively. Note that the PMF has been normalized such that the most stable ion-pairing state is at 0 kJ/mol.

Since the PMF is a concentration-dependent property, we also calculated the PMF at three different low concentrations as shown in Figure 5 (right). Across all concentrations, the main structural features, such as the positions of the CIP (at 3.55 Å) and SSIP (at 5.25 Å) states, as well as the location of the energy barrier separating them (at 4.21 Å), remain consistent. We also note that since the energies of these states are within error bar as the concentration changes, the PMF is effectively converged at 0.4 M concentration (1 $Na_2SO_4$ in 128 water molecules) compared to 0.2 M (1 $Na_2SO_4$ in 256 water molecules), and 0.1 M (1 $Na_2SO_4$ in 512 water molecules) as shown in Figure S16. While some previous studies have predicted higher CIP to SSIP barriers (~12.6 kJ/mol) for other ionic systems such as $CaCl_2$[19] or $NaCl$[3], our results are consistent with more recent studies that predict lower CIP to SSIP barriers (2 - 5 kJ/mol).[9, 24, 94] Discrepancy in the prediction of CIP to SSIP barriers is most likely due to differences in technical details such as the size of the basis set which is important to properly converge forces on the $Na^+$ ions[24] or the nature of the anion. We also observe that at distances above 8 Å at 0.4 M concentration (1 $Na_2SO_4$ in 128 water molecules), there is a significant deviation between the PMF compared to the PMF at 0.2 M and 0.1 M concentrations. This deviation is due to finite size effects as the ions must be in the corners of the simulation box to reach these distances. Consequently, the ions interact unphysically with their periodic images. Therefore, to avoid these errors we further analyze the ion pairing behavior below based on the 2D PMF obtained at 0.1 M concentration.

While the 1D PMF provides some insight into the ion-pairing behavior, it is however limited to interactions involving only one $Na^+$ ion and we lose information regarding the relative abundance of the CIP states ($Na_1SO_4$ or $Na_2SO_4$). Therefore, a 2D PMF is necessary to provide a more comprehensive understanding of the ion-ion and ion-water interactions. Thus, we analyze the 2D PMF obtained from umbrella sampling simulations using 1 $Na_2SO_4$ molecule in 512 water molecules (0.1 M) as shown in Figure 6 below. Here, we observe that having the $Na^+$ ions and the $SO_4^{2-}$ ion fully solvated by water (SSIP in Figure 6, left) is the most energetically stable state. This is in agreement with previous experimental and classical molecular dynamics studies that suggested that at low concentrations, due to strong solvation of the sulfate ion (shown in its high CN), ion-ion interactions are effectively screened thus leading to weaker interaction of the $SO_4^{2-}$ ion with the $Na^+$ ion. [26, 27, 43, 44, 82-84] Using minimum energy path analysis, we observe that there is a 4.5 kJ/mol energy barrier for coordinating the $SO_4^{2-}$ ion with one $Na^+$ ion i.e., the $CIP_1$ state (See Figure S20 for minimum energy pathway) which is 1.6 kJ/mol less stable than the SSIP state (Figure 7, top right). Additionally, there is a 4.4 kJ/mol energy barrier for coordinating the $SO_4^{2-}$ ion with a second $Na^+$ ion ($CIP_2$) which is 1.0 kJ/mol less stable than the $CIP_1$ state, meanwhile it is highly unfavorable to go directly from the SSIP to the fully coordinated $CIP_2$ state as shown the energy barrier of 9.2 kJ/mol (Figure 6, bottom right). Therefore, to place both $Na^+$ ions in direct contact with the $SO_4^{2-}$ ion, we must follow a stepwise approach whereby the cations are coordinated sequentially to the $SO_4^{2-}$ ion (i.e., SSIP -> $CIP_1$ -> $CIP_2$).

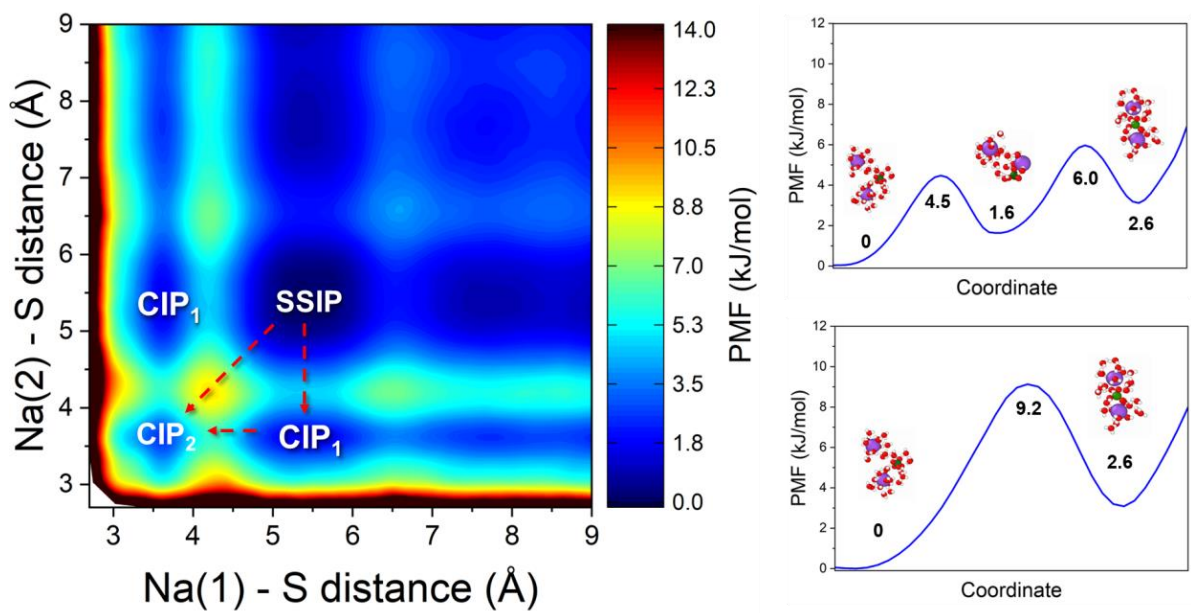

**Figure 6**. 2D PMF of Na$_2$SO$_4$ at 0.4 M concentration using 1 Na$_2$SO$_4$ molecule in 128 water molecules (left), minimum energy path for the SSIP to CIP$_1$ to CIP$_2$ transition (top right), and minimum energy path for the SSIP to CIP$_2$ transition (bottom right). Red arrows represent the minimum energy path between states. SSIP = Solvent-separated ion pair, CIP$_1$ = Na$_1$SO$_4$ contact ion pair, and CIP$_2$ = Na$_2$SO$_4$ contact ion "pair". Off diagonal values have been averaged to smoothen the 2D PMF for visualization purposes, see Figure S18 for raw plot and Figure S19 for the energy difference between the raw and smoothed 2D PMF. Note that the PMF has been normalized such that the most stable ion-pairing state is at 0 kJ/mol. Atom color code: O – red, H – white, Na – purple, S – green.

To further support the above analysis, Figure 7 shows that the SO$_4^{2-}$ ion has strong interactions with water as shown in the S – H (top and bottom left) and sulfate oxygen – H (Os – H, top and bottom right) RDFs (See Table S4 for peak positions, heights and CNs). Here, the first peak of the Os - H RDF (1.82 Å) at 0.1 M concentration corresponds to the hydrogen bonds between the water hydrogen atoms and the oxygens of the sulfate. In addition, the first peak of the S - H RDF is at 2.82 Å at 0.1 M concentration. The close proximity of the hydrogens compared to the Na$^+$ ion (5.25 Å in the SSIP state in the 1D PMF) means that the hydration shell of the SO$_4^{2-}$ ion imposes a steric and energetic barrier as seen in Figure 6 above. Additionally, on average, the sulfate oxygens receive 2.68 HBs from water molecules (at 0.1 M concentration) and thus, to bring the Na$^+$ ion into contact with the SO$_4^{2-}$ ion, the hydrogen bonding network around the sulfate ion must first be broken which would incur an energetic penalty. Therefore, the steric and energetic cost associated with breaking hydrogen bonding network around the SO$_4^{2-}$ ion effectively precludes its close association with Na$^+$ ions and as a result the SO$_4^{2-}$ ion becomes preferentially isolated from the Na⁺ ions. This result is consistent with the previous results that the fully solvated SSIP state should dominate at low concentrations.

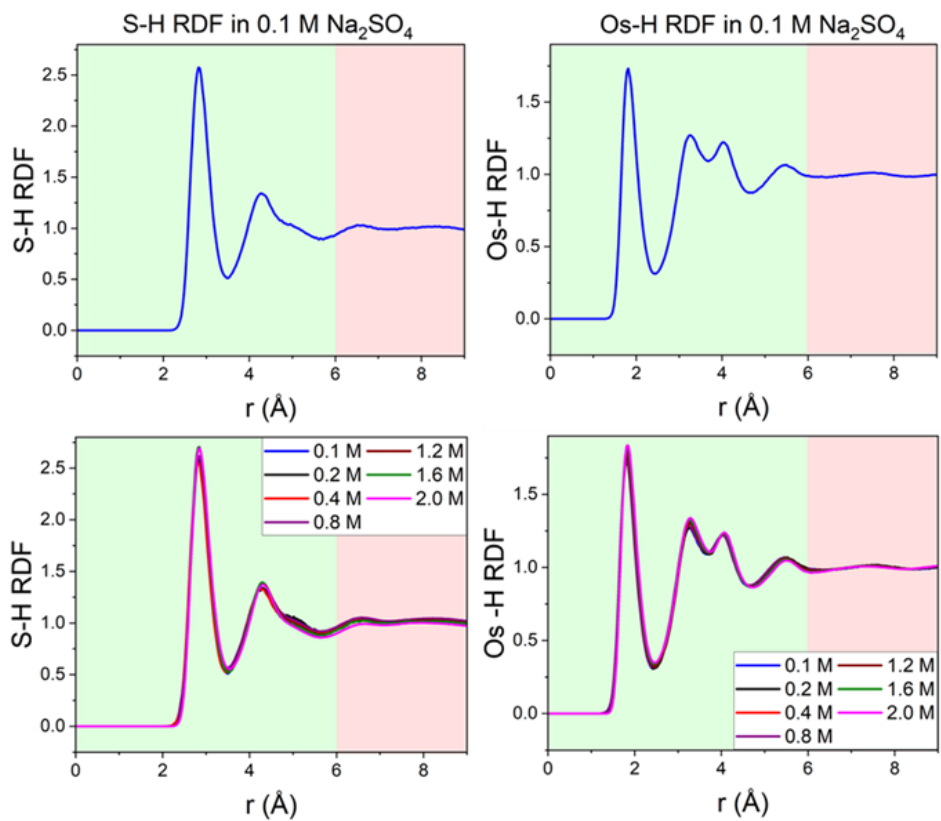

**Figure 7**. Sulfate – hydrogen (top left), sulfate oxygen – hydrogen (top right) RDF at 0.1 M concentration (left), sulfate – hydrogen (bottom left) and sulfate oxygen – hydrogen (bottom right) RDF between 0.1 and 2 M concentration. The green and pink background indicates radial distances that fall within and outside the radial cutoff of our MLIP (6 Å), respectively. Note that the y-axis on plots on the S-H and Os-H plots are on different scales.

**Conclusions**

In this computational study, we demonstrated the feasibility of applying MLIPs developed based on minimal DFT data for studying the structure and behavior of monovalent-divalent ion pairs in water using the example of $Na_2SO_4$ in water. We showed that our trained MLIP can recover density vs concentration trends as well as recover the structural features of bulk water through its RDFs even well beyond the radial cutoff the MLIP (6 Å). We also found that our MLIP reproduces experimentally reported ion hydration structure for the sodium and sulfate ions, respectively. In addition, we also provided the sulfate-sodium RDF which suggested that the ions are preferentially separated from one another and are stabilized by their respective solvation shells. Furthermore, we studied the effect of increasing $Na_2SO_4$ concentration on the local order of bulk water and the diffusivity of water. Here, we observed that increasing $Na_2SO_4$ concentration disrupts the hydrogen bonding network of water and increases disorders, while on the other hand the strong ion-water interactions lead to a decrease in the water diffusivity with concentration. Finally, we explored the free-energy landscape of $Na_2SO_4$ in water through the 1- and 2D PMF. In both cases, our results show that at the low concentration limit, the SSIP state is ~ 3 kJ/mol more stable than the CIP state, thus further reinforcing the notion that the ions are stabilized in their respective solvation shells and are preferentially separated. Minimum energy path analysis of the 2D PMF revealed that to fully coordinate the SO$_4^{2-}$ ion with the Na$^+$ ions, Na$^+$ ions must be coordinated sequentially rather than both at the same time. As a consequence, the SSIP and CIP$_1$ states dominate at the low concentration limit. Overall, we demonstrated the applicability of MLIPs to describe complex aqueous electrolyte solutions that opens up the possibility of developing accurate and reliable MLIPs to further advance the understanding of the intrinsic properties of complex monovalent-divalent electrolyte systems, for example, in acidic or basic conditions, or as a function of temperature. In addition, charge-aware MLIPs may also be used in future work to study properties such as the dielectric constant and dipole moments as a function of temperature or ionic strength, for example. These insights are not only critical for understanding the thermodynamic and kinetic aspects of ion pairing but also have broader implications for processes such as ion transport, reactivity, the design of separation strategies in complex electrolyte systems, and could in part help rationalize the impact of solvated ions at electrochemical interfaces.

**Acknowledgements**

This work was also made possible by the U.S. Department of Energy, Office of Science, CPIMS program, under Award DE-SC0024654. The authors thank Tristan Maxson, Sophia Ezendu, Gbolagade Olajide, Khagendra Baral, Amobi David, and Mustapha Iddrisu for their insightful comments on the manuscript and work. A.S would like to acknowledge the financial support of the University of Alabama Graduate School as a Graduate Council Fellow. This work was also made possible in part by a grant of high-performance computing resources and technical support from the Alabama Supercomputer Authority. This research used computational resources of the National Energy Research Scientific Computing Center (NERSC), a U.S. Department of Energy Office of Science User Facility located at Lawrence Berkeley National Laboratory, operated under Contract DE-AC02-05CH11231 using NERSC Award BES-ERCAP0024218. Any opinions, findings, conclusions, and/or recommendations expressed in this material are those of the authors(s) and do not necessarily reflect the views of funding agencies.

**Supporting Information**

Figures S1–S20, which present parity and error distribution analyses for MLIP-predicted energies, forces, and stresses; distributions of interatomic distances, bond lengths, and coordination numbers in both training and test sets; and comprehensive one- and two-dimensional S–Na PMF results (including convergence, symmetrization, and minimum-energy pathways). Tables S1–S7 summarize bulk water densities, MLIP vs. experimental densities, RDF features, structural parameters, tetrahedral order and hydrogen bonding, and diffusivities of water, sodium, and sulfate as functions of Na$_2$SO$_4$ concentration (PDF).

Raw 1D and 2D S – Na PMF data (EXCEL)

NequIP input yaml files (ZIP)

**TOC Graphic**

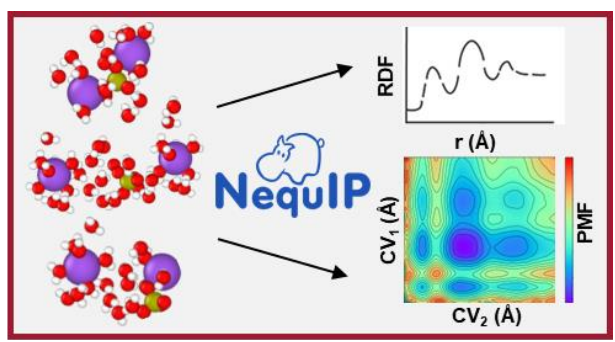